\documentclass[preprint,1p,number]{elsarticle}

\usepackage{multirow}
\usepackage{graphicx}
\usepackage{dcolumn}
\usepackage{bm}
\usepackage{url}                                       
\usepackage{subfigure}
\usepackage{epstopdf}

\journal{Journal of Physics A: Mathematical and Theoretical}

\begin{document}

\begin{frontmatter}

\title{Empirical comparison of network sampling techniques}
\author{Neli Blagus\corref{coraut}}
\ead{neli.blagus@fri.uni-lj.si}
\author{Lovro \v Subelj\corref{}}
\ead{lovro.subelj@fri.uni-lj.si}
\author{Marko Bajec\corref{}}
\ead{marko.bajec@fri.uni-lj.si}
\address{University of Ljubljana, Faculty of Computer and Information Science, Ljubljana, Slovenia}

\begin{abstract}
In the past few years, the storage and analysis of large-scale and fast evolving networks present a great challenge. Therefore, a number of different techniques have been proposed for sampling large networks. In general, network exploration techniques approximate the original networks more accurately than random node and link selection. Yet, link selection with additional subgraph induction step outperforms most other techniques. In this paper, we apply subgraph induction also to random walk and forest-fire sampling. We analyze different real-world networks and the changes of their properties introduced by sampling. We compare several sampling techniques based on the match between the original networks and their sampled variants. The results reveal that the techniques with subgraph induction underestimate the degree and clustering distribution, while overestimate average degree and density of the original networks. Techniques without subgraph induction step exhibit exactly the opposite behavior. Hence, the performance of the sampling techniques from random selection category compared to network exploration sampling does not differ significantly, while clear differences exist between the techniques with subgraph induction step and the ones without it. 

\end{abstract}


\end{frontmatter}

\section{\label{sec:intro}Introduction}
Real-world networks are often very large and fast evolving. Therefore, not only their storage poses a problem, but their analysis and understanding present a great challenge. In the past few years, a number of different techniques have been proposed for sampling large networks to allow for their faster and more efficient analysis. However, some information about the network is lost through sampling, thus it is of key importance to understand the changes in the network structure introduced by sampling.

Several studies on network sampling analyze the match between the original networks and their sampled variants~\cite{HKBG08,SC12,BSWB14}, while only few of them focus on comparing the performance of different sampling techniques~\cite{LF06,LKJ06,BSB14}. In general, network exploration techniques like random walk and forest-fire sampling approximate the original networks more accurately than random node and link selection~\cite{LF06}. However, Ahmed~et~al.~\cite{ANK11} proposed link selection with additional subgraph induction step, where the sampled network consists of randomly selected links (i.e. random link selection) and any additional links between their endpoints (i.e. subgraph induction). In this way, not only the performance of random link selection is improved, the proposed technique also outperforms several other sampling techniques. 

In this paper, we apply subgraph induction also to random walk and forest-fire sampling. We consider ten real-world networks and analyze the changes of their properties introduced by sampling. We compare eight sampling techniques based on the match between the original and sampled networks. The results reveal that techniques with subgraph induction step tend to underestimate the degree and clustering distribution, while other techniques tend to overestimate both properties. Moreover, the techniques with subgraph induction overestimate the average degree and density of the original networks, in contrast, techniques without subgraph induction underestimate them. It appears that the performance of the sampling techniques from random selection category compared to network exploration sampling does not differ significantly, while clear differences exist between the techniques with subgraph induction step and the ones without subgraph induction. 

The rest of the paper is structured as follows. In Section~\ref{sec:sampl}, we first present the background on network sampling and expose the sampling techniques used in the study. The results of the empirical analysis are reported and formally discussed in Section~\ref{sec:analys}, while Section~\ref{sec:conc} concludes the paper and suggests directions for further research.

\section{\label{sec:sampl}Network sampling}
Let the network be represented by a simple undirected graph $G = (V, E)$, where $V$ denotes the set of nodes ($n = |V|$) and $E$ is the set of links ($m = |E|$). The goal of network sampling is to create a sampled network $G' = (V', E')$, where $V' \subset V$, $E' \subset E$ and $n' = |V'| << n$, $m' = |E'| << m$. The sample $G'$ is obtained in two steps. In the first step, nodes or links are sampled using a particular strategy like random selection and network exploration sampling. In the second step, the sampled nodes and links are retrieved from the original network. The sampled network is called a subgraph of the original network, if it consists of sampled nodes or sampled links only. Otherwise, if sampled nodes and all their mutual links are included in the sample or the sample consists of sampled links and any additional links among their endpoints, the sampled network is called an induced subgraph of the original network (i.e., subgraph induction).

However, the previous studies have shown that the size $n'$ of the sample affects the performance of network sampling. Expectedly, larger samples approximate the original networks more accurately~\cite{BSB14}. Still, the sample size of $10-30\%$ of the original network achieves the balance between small sample and accurate approximation~\cite{LF06,DB13,BSB14}.  

The changes of network properties introduced by sampling depend also on the adopted sampling technique, since different techniques are suitable for matching different sets of network properties. Thus, sampling techniques can be roughly divided into two categories: random selection and network exploration techniques. In the first category, nodes or links are included in the sample uniformly at random or proportional to some particular characteristic like degree. In the second category, the sample is constructed by retrieving a neighborhood of a randomly selected seed node using different strategies like breadth-first search, random walk and forest-fire. 

\subsection{\label{subsec:RS}Random selection}
\begin{figure}[!t]
\centering
				\subfigure[\ RNS\label{subfig:RNS}]{\includegraphics[width=0.18\columnwidth]{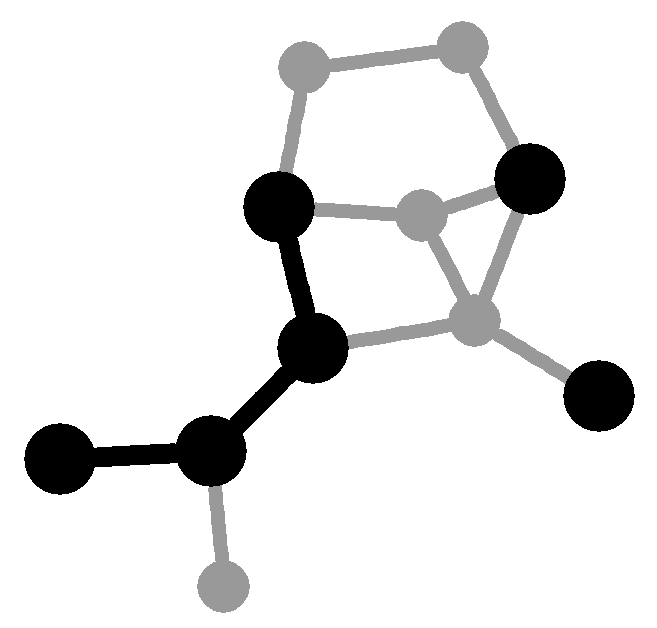}}\quad\quad
				\subfigure[\ RND\label{subfig:RND}]{\includegraphics[width=0.18\columnwidth]{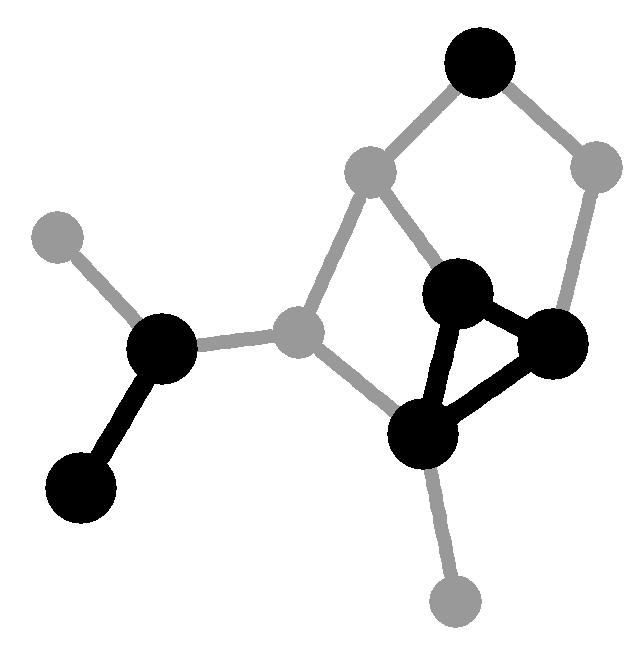}}\quad\quad
				\subfigure[\ RLS\label{subfig:RLS}]{\includegraphics[width=0.18\columnwidth]{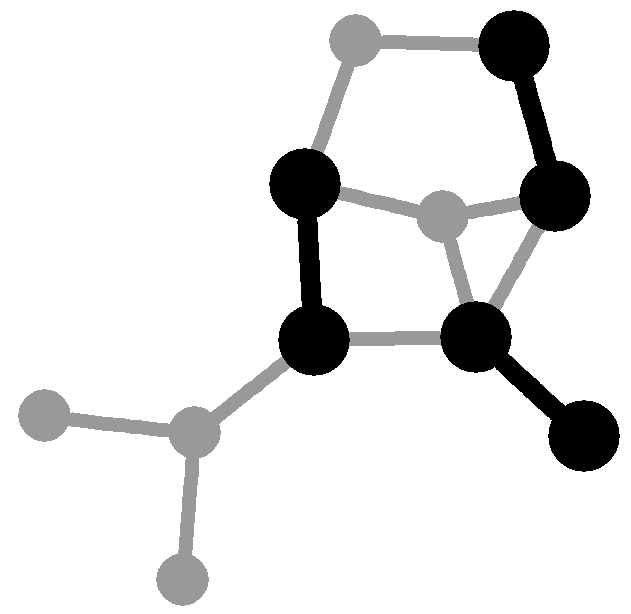}}\quad\quad
				\subfigure[\ RLI\label{subfig:RLI}]{\includegraphics[width=0.18\columnwidth]{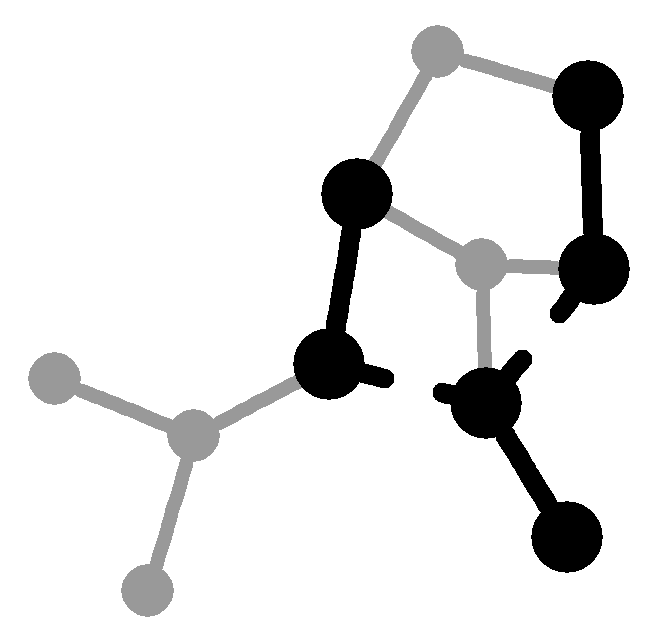}} 
				\caption{Random selection techniques applied to a small toy network. Highlighted nodes and links represent the samples obtained by different techniques. \subref{subfig:RNS} In random node selection, the sample consists of nodes selected uniformly at random and all their mutual links. \subref{subfig:RND} In random node selection by degree, the nodes are selected to the sample with probability proportional to their degree, while all their mutual links are included in the sample. \subref{subfig:RLS} In random link selection, links are selected to the sample uniformly at random. \subref{subfig:RLI} In random link selection with subgraph induction, the sample consists of randomly selected links (solid lines) and also any additional links between their endpoints (dashed lines).}
\label{fig:RS}
\end{figure}

For the purpose of this study, we consider four techniques from the random selection category. We first adopt random node selection~\cite{LF06} (RNS), where the sample consists of nodes selected uniformly at random and all their mutual links (Fig.~\ref{subfig:RNS}). RNS accurately approximates the degree mixing ~\cite{LKJ06} and preserves the relationship of transitivity and density between the original and sampled networks~\cite{BSB14}. Moreover, it shows better performance on larger samples than on smaller~\cite{BSB14}. Yet, RNS overestimates the degree and betweenness centrality exponent and fails to match the clustering coefficient~\cite{LKJ06}, degree distribution~\cite{SWM05} and the average path length~\cite{ANK10} of the original network. 

Furthermore, we adopt random node selection by degree~\cite{LF06} (RND), which improves the performance of RNS. Here, the nodes are selected randomly with probability proportional to their degrees and all their mutual links are included in the sample (Fig.~\ref{subfig:RND}). RND matches in-degree and out-degree distributions and also spectral properties of the original network better than RNS~\cite{LF06}. Besides, it constructs samples with larger weakly connected component~\cite{BSB14}. Nevertheless, despite a fully connected original network, both RNS and RND can construct a disconnected sampled network.

Next, we adopt random link selection~\cite{LF06} (RLS), where the sample consists of links selected uniformly at random (Fig.~\ref{subfig:RLS}). RLS matches well degree mixing~\cite{LKJ06} and the distribution of sizes of weakly connected components~\cite{LF06}. It constructs sparse samples and accurately approximates the average path length of the original network~\cite{ANK12}. Yet, RLS fails to match most of other network properties~\cite{LF06}. Like RNS, RLS overestimates the degree and betweenness centrality exponent and underestimates the clustering coefficient~\cite{LKJ06}. 

We last adopt random link selection with subgraph induction~\cite{ANK11} (RLI), which improves the performance of RLS. Here, the sample consists of links selected uniformly at random and any additional links between their endpoints (Fig.~\ref{subfig:RLI}). RLI outperforms several other techniques in matching the degree, path length and clustering coefficient distribution of the original networks~\cite{ANK11}. It selects nodes with higher degree more likely than other random selection techniques, which increases the connectivity of the sample. Moreover, RLI is suitable for sampling large networks that can not fit into the main memory and can also be implemented as a technique for sampling streaming networks~\cite{ANK12}. 

\subsection{\label{subsec:NE}Network exploration}
\begin{figure}[!t]
\centering
				\subfigure[\ RWS\label{subfig:RWS}]{\includegraphics[width=0.18\columnwidth]{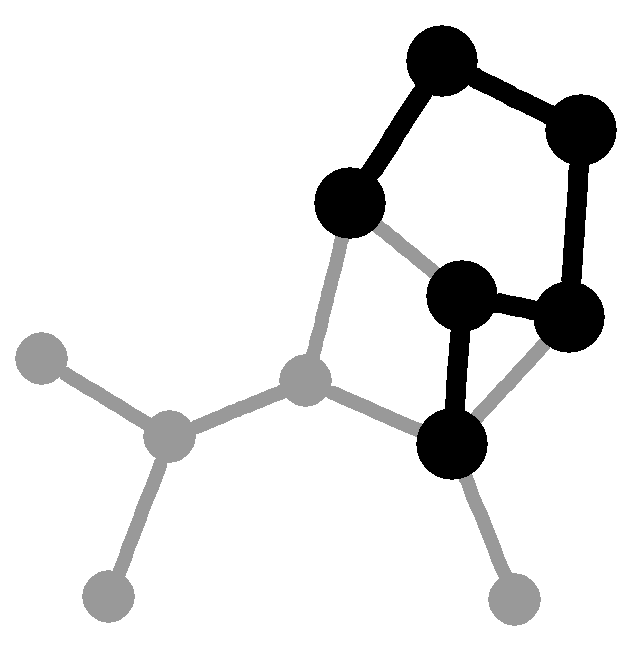}}\quad\quad
				\subfigure[\ RWI\label{subfig:RWI}]{\includegraphics[width=0.18\columnwidth]{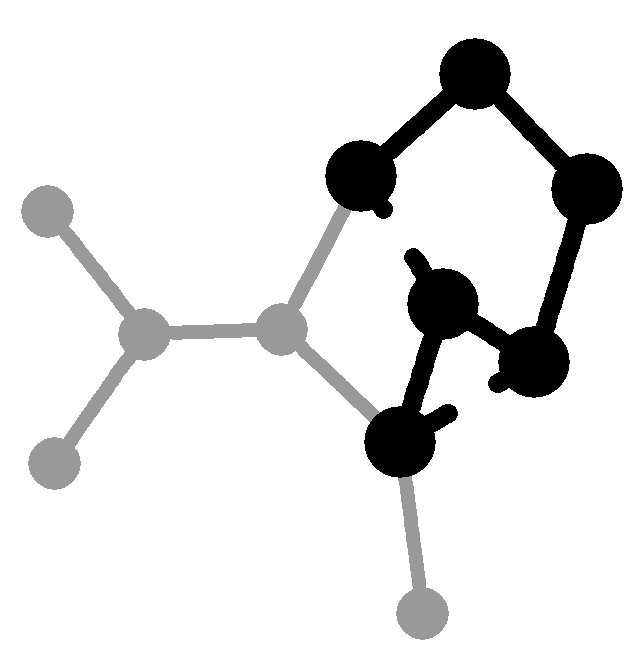}}\quad\quad
				\subfigure[\ FFS\label{subfig:FFS}]{\includegraphics[width=0.18\columnwidth]{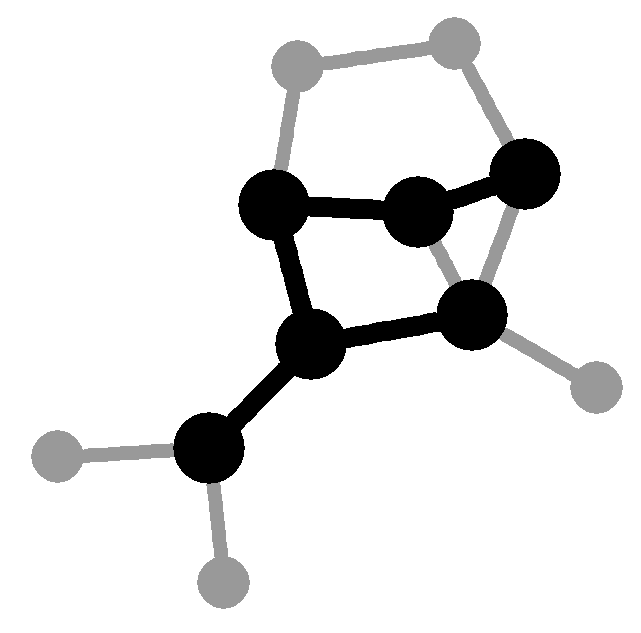}}\quad\quad
				\subfigure[\ FFI\label{subfig:FFI}]{\includegraphics[width=0.18\columnwidth]{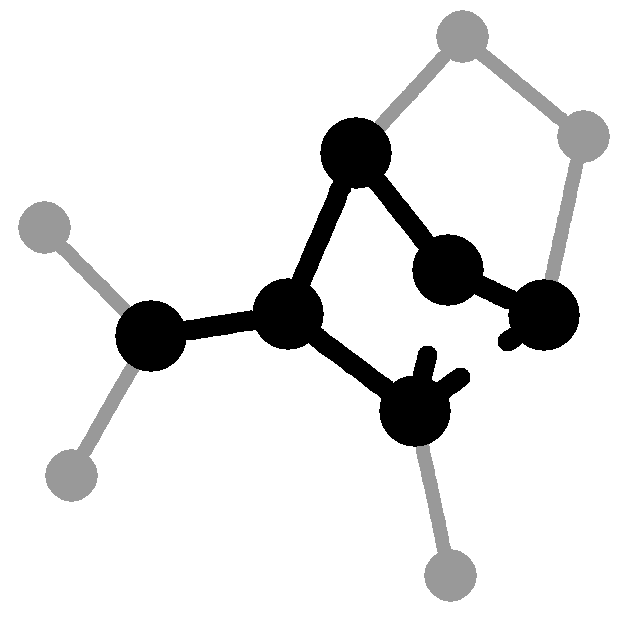}} 
				\caption{Network exploration techniques applied to a small toy network. Highlighted nodes and links represent the samples obtained by different techniques. \subref{subfig:RWS} In random walk sampling, the sample consists of links, retrieved from a simulation of a random walker on the network starting at a randomly selected seed node. \subref{subfig:RWI} In random walk sampling with subgraph induction, the sample consists of links selected with random walk sampling (solid lines) and also any additional links between their endpoints (dashed lines). \subref{subfig:FFS} In forest-fire sampling, the broad neighborhood of a randomly selected seed node is retrieved using partial breadth-first search, where only a fraction of links is included in the sample on each step. \subref{subfig:FFI} In forest-fire sampling with subgraph induction, the sample consists of links selected with forest-fire sampling (solid lines) and also any additional links between their endpoints (dashed lines).}
\label{fig:NE}
\end{figure}

We consider four sampling techniques from the network exploration category (note that in the literature, this category of sampling techniques is also called topology based sampling~\cite{ANK11}, traversal based sampling~\cite{HL13} or link-trace sampling~\cite{MBT11}). First, we adopt random walk sampling~\cite{LF06} (RWS), where the random walk is simulated on the network, starting at a randomly selected seed node (Fig.~\ref{subfig:RWS}). The sample consists of links, which are visited by a random walker and represents a connected subgraph of the original network. RWS outperforms random selection techniques in matching the transitivity~\cite{MBT11}, clustering coefficient distribution and spectral properties and also shows good performance on smaller samples~\cite{LF06}. Yet, RWS is biased towards selecting nodes with high degree~\cite{lovas93} and fails to match the degree distribution~\cite{KMT10}.

Next, we adopt forest-fire sampling~\cite{LF06} (FFS). Here, a broad neighborhood of a randomly selected seed node is retrieved from partial breadth-first search (Fig.~\ref{subfig:FFS}). The number of links sampled on each step is selected from a geometric distribution with mean $p/(1-p)$, where $p$ is set to $0.7$~\cite{LF06}. Thus, on average $2.33$ links are included in the sample on each step. FFS matches well spectral properties~\cite{LF06} and together with RWS shows the best overall performance among several techniques~\cite{LF06}. However, FFS fails to match the path length and clustering coefficient of the original networks~\cite{ANK12}. 

Moreover, we apply subgraph induction step to random walk and forest-fire sampling, which we term random walk sampling with subgraph induction (RWI) and forest-fire sampling with subgraph induction (FFI). Here, the sample consists of links, sampled with random walk (Fig.~\ref{subfig:RWI}) or forest-fire sampling (Fig.~\ref{subfig:FFI}), while any additional links among the endpoints of sampled links are also included in the sample. To the best of our knowledge, RWI has not been analysed in any of the previous studies. On the other hand, FFI shows worse performance than RLI in matching the path length, degree and clustering distributions~\cite{ANK12}. Still, the performance of FFI has not yet been compared to a larger set of sampling techniques.

\section{\label{sec:analys}Analysis and discussion}
In the following sections, we present the adopted framework for comparison of sampling techniques (Section~\ref{subsec:stats}), report the results of the empirical analysis and formally discuss the findings (Section~\ref{subsec:analysis}).

\subsection{\label{subsec:stats}Statistical comparison}
Sampling techniques are compared through four network properties (see Section~\ref{subsec:analysis}) on ten real-world networks (see Table~\ref{tbl:nets}). For each property, we compute externally studentized residuals of the properties of sampled networks that measure the consistency of each sampling technique with the rest. We expose statistically significant inconsistencies between the techniques with two-tailed Student $t-$test~\cite{CW82} and calculate the residuals as introduced below (see also~\cite{SFB14}).

Let $x_{ij}$ denote some property value of $j-$th network for $i-$th sampling technique. Externally studentized residual $\hat{x}_{ij}$ is calculated as

\begin{equation} 
\label{eq:1}
\hat{x}_{ij} = \frac{x_{ij}-\hat{\mu}_{ij}}{\hat{\sigma}_{ij}\sqrt{1-\frac{1}{N}}},
\end{equation}

where $N$ is the number of sampling techniques ($N = 8$ in our case), $\hat{\mu}_{ij}$ is the mean value and $\hat{\sigma}_{ij}$ is the standard deviation of the considered property. The mean value and standard deviation are computed for all sampling techniques, excluding the observed one

\begin{equation}
\hat{\mu}_{ij} = \sum_{k \neq i}{\frac{x_{kj}}{(N-1)}},
\end{equation}

\begin{equation}
\label{eq:3}
\hat{\sigma}_{ij}^{2} = \sum_{k \neq i}{\frac{(x_{kj} - \hat{\mu}_{ij})^2}{(N-2)}}.
\end{equation}

Assuming that the errors in $x$ are independent and normally distributed, the residuals $\hat{x}$ have Student $t-$distribution with $N-2$ degrees of freedom. Statistically significant inconsistencies between the sampling techniques are revealed by two-tailed Student $t-$test~\cite{CW82} at $P-$value of $0.05$, rejecting the null hypothesis that the values of the considered property are consistent across the sampling techniques.

\subsection{\label{subsec:analysis}Empirical analysis}
The empirical analysis is performed on ten social and information networks. Their main characteristics are presented in Table~\ref{tbl:nets}. Due to a large number of networks considered, the detailed description is omitted. Networks are considered to be undirected, although some of them are directed. For each network, we perform $100$ realizations of each sampling technique, where we consider sample sizes of $15\%$ of the original networks as suggested in~\cite{LF06,BSB14}. For each run of the network exploration techniques, the sample was constructed from a new randomly selected seed node. 

We analyze different properties of the original and sampled networks, including degree distribution (probability distribution of degrees of all nodes), distribution of clustering coefficient (probability distribution of the proportions of connected neighbors of each node~\cite{WS98}), average degree (average number of neighbors of nodes over the whole network), and density (the ratio of existing links to all possible links). We compare sampling techniques based on the match between the original and sampled networks; Section~\ref{subsec:deg} reports the results for degree and clustering distributions, while in Section~\ref{subsec:clus} we analyze the average degree and density.

\begin{table}[t]
\scriptsize
\centering
\caption{\label{tbl:nets}Real-world networks considered in the study.}
\begin{tabular}{clrrccc}
\hline\noalign{\smallskip}
\multicolumn{1}{c}{\multirow{2}{*}{Network}} & \multicolumn{1}{c}{\multirow{2}{*}{Description}} & \multicolumn{1}{c}{\multirow{2}{*}{Nodes}} & \multicolumn{1}{c}{\multirow{2}{*}{Links}} & \multicolumn{1}{c}{Average} & \multicolumn{1}{c}{Clustering} & \multicolumn{1}{c}{\multirow{2}{*}{Density}} \\
& & & & \multicolumn{1}{c}{degree} & \multicolumn{1}{c}{coefficient} & 
\\\noalign{\smallskip}\hline\noalign{\smallskip}
\textit{ca-hep} & \textit{High E. Phys.} collaboration~\cite{LKF07} &	$12$,$008$ &	$237$,$010$ & $39.5$ & $0.660$ & $3.3 \times 10^{-3}$ \\
\textit{ca-astro} & \textit{Astro Phys.} collaboration~\cite{LKF07} &	$18$,$772$ &	$396$,$160$ & $42.2$ & $0.318$ & $2.2 \times 10^{-3}$ \\
\textit{cit-hep} & \textit{High E. Phys.} citation~\cite{KDD} &	$27$,$240$ &	$342$,$437$ & $25.1$ & $0.120$ & $9.2 \times 10^{-4}$ \\
\textit{brightkite} & \textit{Brightkite} friendship~\cite{CML11}  &	$58$,$228$ &	$214$,$078$ & $\phantom{4}7.4$ & $0.111$ & $1.3 \times 10^{-4}$ \\
\textit{slashdot} & \textit{Slashdot} friendship~\cite{LLDM09} &	$82$,$168$ &	$948$,$464$ & $23.1$ & $0.024$ & $2.8 \times 10^{-4}$ \\
\textit{flickr} & \textit{Flickr} images metadata~\cite{ML12} &	$105$,$938$ &	$2$,$316$,$948$ & $43.7$ & $0.402$ & $4.1 \times 10^{-4}$ \\
\textit{ca-dblp} & DBLP collaboration~\cite{YL12} &	$317$,$080$ &	$1$,$049$,$866$ & $\phantom{4}6.6$ & $0.306$ & $2.1 \times 10^{-5}$ \\
\textit{nd.edu} & Web graph of \textit{nd.edu}~\cite{AJB99} &	$325$,$729$ &	$1$,$497$,$134$ & $\phantom{4}9.1$ & $0.097$ & $2.8 \times 10^{-5}$ \\
\textit{youtube} & \textit{Youtube} friendship~\cite{YL15} &	$1$,$134$,$890$ &	$2$,$987$,$624$ & $\phantom{4}5.2$ & $0.006$ & $4.5 \times 10^{-6}$ \\
\textit{road-tx} & Texas road network~\cite{LLDM09} &	$1$,$379$,$917$ &	$1$,$921$,$660$ & $\phantom{4}5.5$ & $0.060$ & $3.9 \times 10^{-6}$ \\
\noalign{\smallskip}\hline
\end{tabular}
\end{table}

\subsubsection{\label{subsec:deg}Degree and clustering coefficient distributions}
\begin{figure}[ht]
\centering
\subfigure[Degree distribution]{\includegraphics[width=0.75\columnwidth]{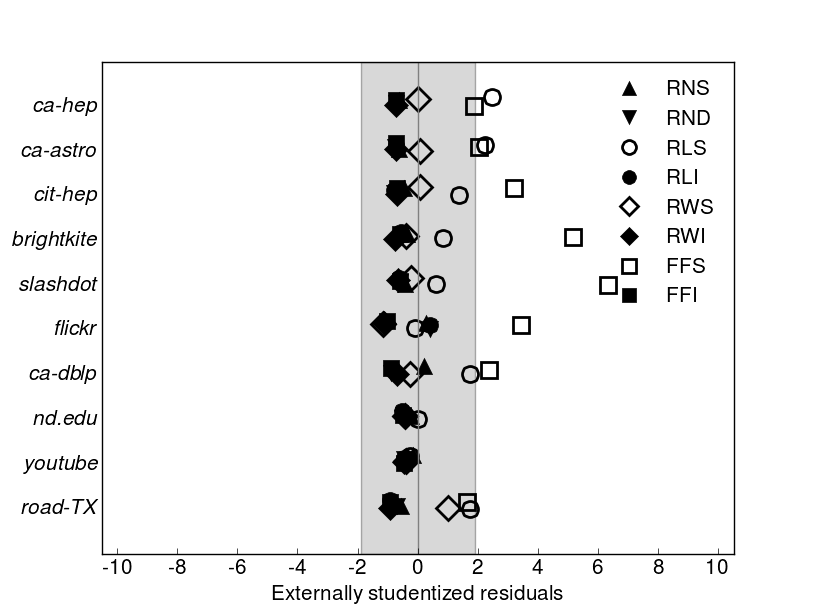}\label{subfig:deg-all}} \quad
\subfigure[Degree distribution]{\includegraphics[width=0.75\columnwidth]{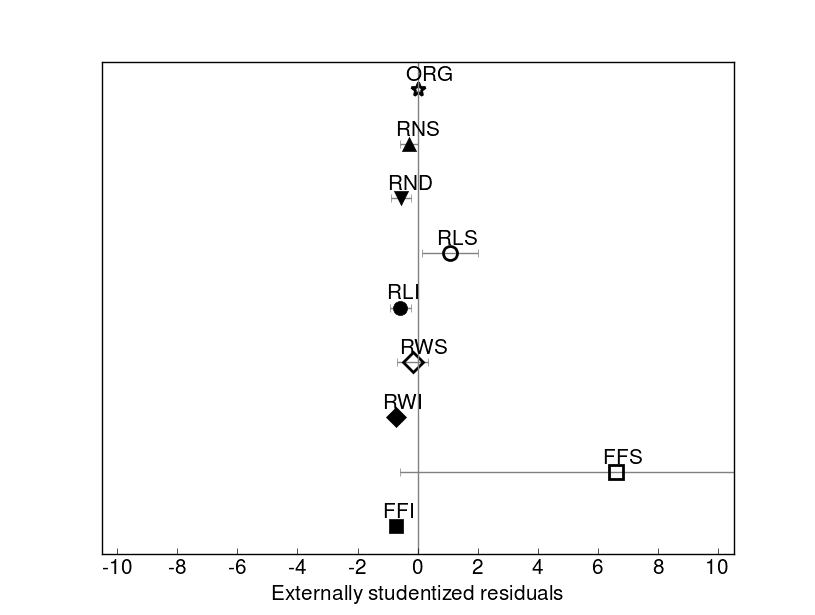}\label{subfig:deg-avg}} \caption{Statistical comparison of the degree distribution of the original networks and their sampled variants obtained by different sampling techniques. We show externally studentized residuals that measure the consistency of each sampling technique with the rest and expose statistically significant inconsistencies between the techniques with two tailed Student $t-$test at $P-$value of $0.05$ (see~\subref{subfig:deg-all}; shaded region corresponds to $95\%$ confidence intervals). Merely for complete understanding, we present average externally studentized residuals over all networks (see~\subref{subfig:deg-avg}; the star marker indicate the residuals for the original networks, while error bars display standard deviations). Notice that techniques with subgrah induction step (full markers) approximate the degree distribution more accurately than the techniques without subgraph induction (empty markers).}
\label{fig:degdist}
\end{figure}

\begin{figure}[ht]
\centering
\subfigure[Clustering coefficient distribution]{\includegraphics[width=0.75\columnwidth]{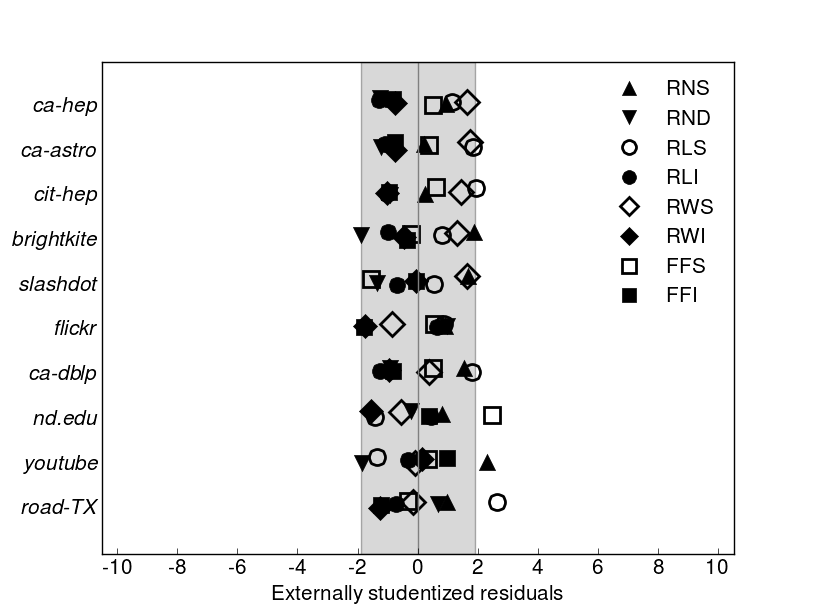}\label{subfig:clus-all}} \quad
\subfigure[Clustering coefficient distribution]{\includegraphics[width=0.75\columnwidth]{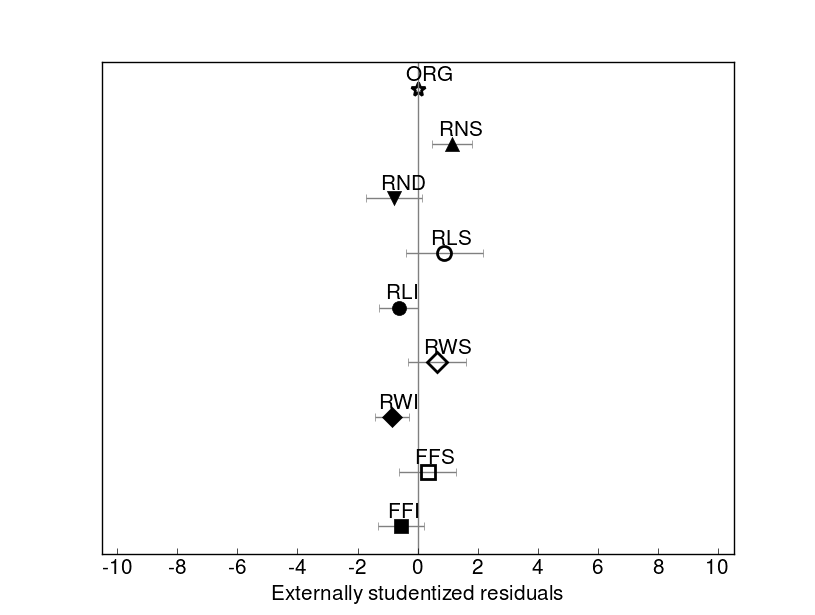}\label{subfig:clus-avg}} \caption{Statistical comparison of the clustering coefficient distribution of the original networks and their sampled variants obtained by different sampling techniques. We show externally studentized residuals that measure the consistency of each sampling technique with the rest and expose statistically significant inconsistencies between the techniques with two tailed Student $t-$test at $P-$value of $0.05$ (see~\subref{subfig:clus-all}; shaded region corresponds to $95\%$ confidence intervals). Merely for complete understanding, we present average externally studentized residuals over all networks (see~\subref{subfig:clus-avg}; the star marker indicate the residuals for the original networks, while error bars display standard deviations). Notice that techniques with subgrah induction step (full markers) improve the performance of the corresponding techniques without subgraph induction (empty markers).}
\label{fig:clus}
\end{figure}

We first analyze the performance of sampling techniques based on the match of the degree and clustering coefficient distributions between the original and sampled networks. To compare the distributions of properties, we use Kolmogorov-Smirnov $D-$statistics, which is commonly used in similar studies~\cite{LF06,ANK12,BSB14}. Kolmogorov-Smirnov test checks the null hypothesis that the distributions of property of the original network and its sampled variant are the same, while the $D-$statistics measures the distance between the observed distributions. Based on the values of Kolmogorov-Smirnov $D-$statistics, we compare the adopted sampling techniques (see Section~\ref{subsec:stats}).

The comparison of sampling techniques based on the degree distribution is shown in Fig.~\ref{fig:degdist}. We observe that in most cases, the difference between the techniques with subgraph induction step (i.e., RNS, RND, RLI, RWI, FFI) and those without subgraph induction (i.e., RLS, RWS, FFS) is clear. The first group of techniques approximates the degree distribution of the original networks more accurately. In addition, the techniques with subgraph induction step improve the performance of the corresponding techniques without subgraph induction in the case of RLS and FFS. Since the subgraph induction increases the degrees of the nodes in the sample~\cite{ANK11}, this clearly contributes to better match the degree distribution between the original and sampled networks. 

RNS and RND show comparable accuracy to network exploration techniques with subgraph induction, even though they construct disconnected samples. The latter indicates that the connectivity of the sample does not affect the match of the degree distribution. On the other hand, among the techniques without subgraph induction, RWS shows the best performance, which could be explained by its bias towards selecting high degree nodes and exploring densely connected parts of the network~\cite{LF06}. In contrast, the samples constructed by FFS are sparse, a large fraction of the nodes in the samples has low degree, while the number of nodes with higher degree is underestimated~\cite{ANK11}. Accordingly, FFS is the least accurate among all techniques.

Moreover, we compare the sampling techniques based on the clustering coefficient distribution. The results are presented in Fig.~\ref{fig:clus}. The techniques show more comparable accuracy than in the case of the degree distribution. In general, the techniques with subgraph induction underestimates the clustering distribution, while others overestimate it. FFS and RWS prove as the best performing techniques. This indicates that in the case of matching the clustering distribution of the original networks, including all the links among sampled nodes is not the best choice. However, RNS shows the worst performance, which could be explained by its tendency to construct samples with a large number of low clustered nodes~\cite{ANK11}. 

In summary, the sampling techniques with subgraph induction step underestimate the degree and clustering coefficient distributions of the original networks. On the other hand, the techniques without subgraph induction, overestimates both properties. The best overall performance is provided by RWI, which shows the most stable accuracy for matching both distributions on all considered networks. However, it appears that the properties of the original networks have relevant effect on the accuracy of the sampling techniques. For example, the techniques with subgraph induction match the degree distribution of the networks with larger average degree more accurately than when the average degree is lower and the techniques without subgraph induction also perform well. We aim to study that kind of relations in future work, since they should be observed in an even larger set of real-world networks to obtain statistically significant results.

\subsubsection{\label{subsec:clus}Average degree and density}
In the second part of the analysis, we study the performance of sampling techniques based on the match of the average degree and density between the original and sampled networks. We compare the properties between networks based on the actual values of properties. Furthermore, we modify Eq.~(\ref{eq:1}) and~(\ref{eq:3}) and use the true value of considered property of the original networks instead of the mean values $\hat{\mu}_{ij}$ (see Section~\ref{subsec:stats}) to compare the sampling techniques.

\begin{figure}[!t]
\centering
\subfigure[Average degree]{\includegraphics[width=0.75\columnwidth]{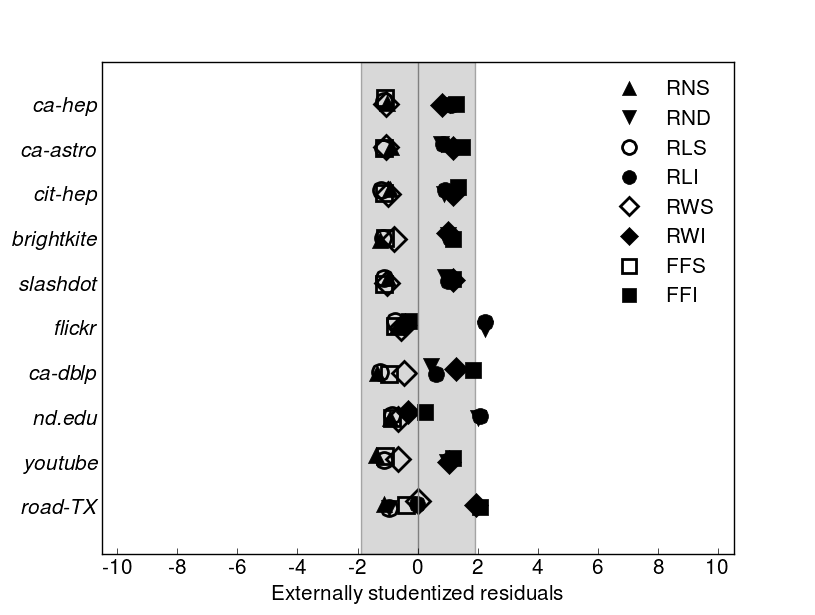}\label{subfig:avgdeg-all}} \quad
\subfigure[Average degree]{\includegraphics[width=0.75\columnwidth]{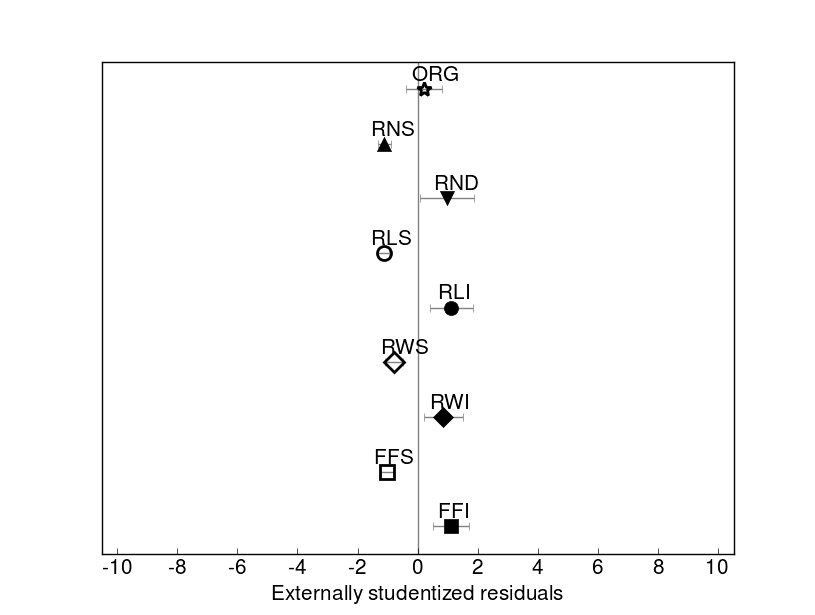}\label{subfig:avgdeg-avg}} \caption{Statistical comparison of the average degree of the original networks and their sampled variants obtained by different sampling techniques. We show externally studentized residuals that measure the consistency of each sampling technique with the rest and expose statistically significant inconsistencies between the techniques with two tailed Student $t-$test at $P-$value of $0.05$ (see~\subref{subfig:avgdeg-all}; shaded region corresponds to $95\%$ confidence intervals). Merely for complete understanding, we present average externally studentized residuals over all networks (see~\subref{subfig:avgdeg-avg}; the star marker indicates the residuals for the original networks, while error bars display standard deviations). Notice that techniques with subgrah induction step (full markers) tend to overestimate the average degree, while the techniques without subgraph induction (empty markers) underestimate it.}
\label{fig:avgdeg}
\end{figure}

Fig.~\ref{fig:avgdeg} shows the comparison of sampling techniques based on the average degree. We observe that the techniques with subgraph induction tend to overestimate the average degree, while others underestimate it. Among the techniques with induction step, only RNS underestimate the average degree. The samples constructed with RNS consist of a large fraction of low-degree nodes~\cite{ANK11} and have a low average degree particularly in the smaller samples~\cite{LKJ06}. Therefore, RNS underestimates the average degree of the networks, especially compared to the techniques, which are biased in selecting nodes with higher degree (i.e., RND).

\begin{figure}[!t]
\centering
\subfigure[Density]{\includegraphics[width=0.75\columnwidth]{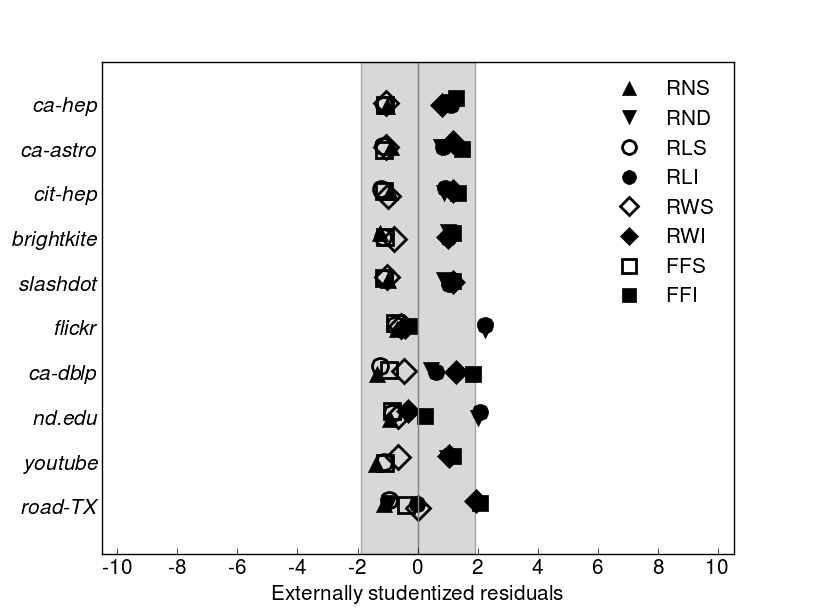}\label{subfig:dens-all}} \quad
\subfigure[Density]{\includegraphics[width=0.75\columnwidth]{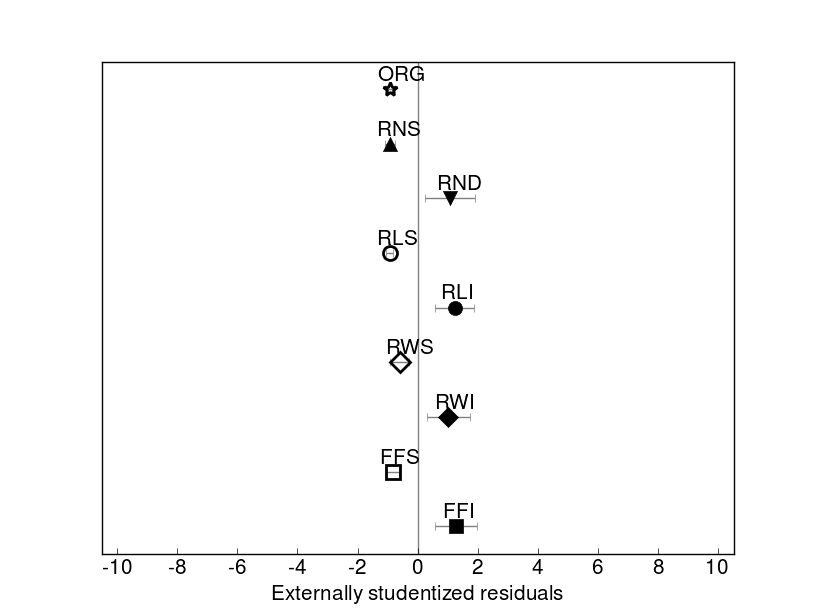}\label{subfig:dens-avg}} \caption{Statistical comparison of the density of the original networks and their sampled variants obtained by different sampling techniques. We show externally studentized residuals that measure the consistency of each sampling technique with the rest and expose statistically significant inconsistencies between the techniques with two tailed Student $t-$test at $P-$value of $0.05$ (see~\subref{subfig:dens-all}; shaded region corresponds to $95\%$ confidence intervals). Merely for complete understanding, we present average externally studentized residuals over all networks (see~\subref{subfig:dens-avg}; the star marker indicates the residuals for the original networks, while error bars display standard deviations). Notice that the techniques without subgraph induction step (empty markers) approximate the density more accurately than the corresponding techniques with subgraph induction (full markers).}
\label{fig:dens}
\end{figure}

Lastly, we analyze the density of the networks. Fig.~\ref{fig:dens} shows the comparison of the sampling techniques based on the density. In general, all techniques overestimate the density, yet the techniques without subgraph induction step approximate the density more accurately than the techniques with subgraph induction. Additionally, techniques without subgraph induction improve the performance of the corresponding techniques with subgraph induction. In the previous study~\cite{BSB12}, we proved the power-law relationship between the size and density of real-world networks and their sampled variants. The results show that the network density decreases with its size. However, the techniques with subgraph induction do follow the latter relationship, while the techniques without induction construct sparser samples than expected. Therefore, the accuracy of the sampling techniques based on the density is relative and depends on whether the samples should accurately match the density of the original networks or the relationship between the size and density of the original networks and their sampled variants should be preserved.

In addition to all above, we analyze the performance of the techniques with partial subgraph induction, where we include a different portion of mutual links between the sampled nodes in the sample (i.e., we randomly select $10\%-90\%$ links from all possible links). The results are not described, since the techniques with partial subgraph induction did not improve the performance of the techniques with subgraph induction and they did not perform worse than techniques without subgraph induction.

In summary, the sampling techniques with subgraph induction step tend to overestimate the average degree and the density of the original networks. In contrast, the techniques without subgraph induction underestimate the average degree and approximate the density of the original networks more accurately than other techniques. Yet, the accuracy of sampling techniques does not depend only on the characteristics of the adopted technique, but also on the characteristics of the original networks. However, from exploration sampling techniques, the best performance is provided by RWI, while the best overall performance is provided by RNS, which shows very stable accuracy in matching considered properties. 

\section{\label{sec:conc}Conclusion}
In this paper, we analyze different real-world networks and study the changes of their properties introduced by network sampling. We consider eight sampling techniques with and without subgraph induction step and compare them based on the match of properties between the original networks and their sampled variants. 

The results reveal that the techniques with subgraph induction underestimate the degree and clustering distribution, while they overestimate the average degree and density of the original networks. On the other hand, the techniques without subgraph induction step exhibit opposite behavior, since they overestimate the degree and clustering distribution and underestimate the average degree and density. The techniques without subgraph induction approximate the density most accurately. However, from the random selection category, random node selection shows the best overall performance, while from the network exploration category, the most stable accuracy is provided by random walk sampling with subgraph induction. Still, it appears that the performance of the sampling techniques from random selection category compared to network exploration sampling does not differ significantly, while clear differences exist between the techniques with subgraph induction step and the ones without subgraph induction. Following this, a suitable classification of network sampling techniques would be a class of techniques with subgraph induction and a class without it. 

Some of the results suggest that the characteristics of the original networks affect the accuracy of the network sampling. Thus, our future work will mainly focus on analyzing relations between network properties and accuracy of the sampling techniques on a larger set of real-world networks. Besides, a prominent direction for further study is the analysis of the time and space efficiency of sampling techniques, since, for example, fitting even a sampled network in the main memory becomes challenging with significant growth of real-world networks in the past few years.



\section*{Acknowledgment}
This work has been supported in part by the Slovenian Research Agency \textit{ARRS} within the Research Program No. P2-0359, by the Slovenian Ministry of Education, Science and Sport Grant No. 430-168/2013/91, and by the European Union, European Social Fund.


\section*{References}
\bibliographystyle{elsarticle-num}

\begin{thebibliography}{10}
\expandafter\ifx\csname url\endcsname\relax
  \def\url#1{\texttt{#1}}\fi
\expandafter\ifx\csname urlprefix\endcsname\relax\def\urlprefix{URL }\fi
\expandafter\ifx\csname href\endcsname\relax
  \def\href#1#2{#2} \def\path#1{#1}\fi

\bibitem{HKBG08}
C.~H{\"u}bler, H.~P. Kriegel, K.~Borgwardt, Z.~Ghahramani, Metropolis
  algorithms for representative subgraph sampling, in: Proceedings of the 8th
  International Conference on Data Mining, IEEE, 2008, pp. 283--292.

\bibitem{SC12}
H.~Sethu, X.~Chu, A new algorithm for extracting a small representative
  subgraph from a very large graph, e-print {arXiv:1207.4825}.

\bibitem{BSWB14}
N.~Blagus, L.~{\v{S}}ubelj, G.~Weiss, M.~Bajec, Sampling promotes community
  structure in social and information networks, Physica A 432 (2015) 206--215.

\bibitem{LF06}
J.~Leskovec, C.~Faloutsos, Sampling from large graphs, in: Proceedings of the
  12th ACM SIGKDD International Conference on Knowledge Discovery and Data
  Mining, ACM, 2006, pp. 631--636.

\bibitem{LKJ06}
S.~H. Lee, P.~J. Kim, H.~Jeong, Statistical properties of sampled networks,
  Phys. Rev. E 73~(1) (2006) 016102.

\bibitem{BSB14}
N.~Blagus, L.~{\v{S}}ubelj, M.~Bajec, Assessing the effectiveness of real-world
  network simplification, Physica A 413 (2014) 134--146.

\bibitem{ANK11}
N.~Ahmed, J.~Neville, R.~R. Kompella, Network sampling via edge-based node
  selection with graph induction, Tech. rep., Purdue University (2011).

\bibitem{DB13}
C.~Doerr, N.~Blenn, Metric convergence in social network sampling, in:
  Proceedings of the 5th ACM workshop on HotPlanet, ACM, 2013, pp. 45--50.

\bibitem{SWM05}
M.~P.~H. Stumpf, C.~Wiuf, R.~M. May, Subnets of scale-free networks are not
  scale-free: sampling properties of networks, P. Natl. Acad. Sci. USA 102~(12)
  (2005) 4221--4224.

\bibitem{ANK10}
N.~K. Ahmed, J.~Neville, R.~Kompella, Reconsidering the foundations of network
  sampling, in: Proceedings of the 2nd Workshop on Information in Networks,
  2010.

\bibitem{ANK12}
N.~K. Ahmed, J.~Neville, R.~Kompella, Network sampling: from static to
  streaming graphs, e-print {arXiv:11211.3412}.

\bibitem{HL13}
P.~Hu, W.~C. Lau, A survey and taxonomy of graph sampling, e-print
  {arXiv:1308.5865}.

\bibitem{MBT11}
A.~S. Maiya, T.~Y. Berger-Wolf, Benefits of bias: towards better
  characterization of network sampling, in: Proceedings of the 17th ACM SIGKDD
  International Conference on Knowledge Discovery and Data Mining, ACM, 2011,
  pp. 105--113.

\bibitem{lovas93}
L.~Lov{\'a}sz, Random walks on graphs: A survey, Combinatorics: Paul Erd{\"o}s
  is eighty 2~(1) (1993) 1--46.

\bibitem{KMT10}
M.~Kurant, A.~Markopoulou, P.~Thiran, On the bias of {BFS}, in: Proceedings of
  the 22nd International Teletraffic Congress, IEEE, 2010, pp. 1--8.

\bibitem{CW82}
R.~D. Cook, S.~Weisberg, Residuals and influence in regression, New York:
  Chapman and Hall, 1982.

\bibitem{SFB14}
L.~{\v{S}}ubelj, D.~Fiala, M.~Bajec, Network-based statistical comparison of
  citation topology of bibliographic databases, Sci. Rep. 4 (2014) 6496.

\bibitem{WS98}
D.~J. Watts, S.~H. Strogatz, Collective dynamics of ''small-world'' networks,
  Nature 393~(6684) (1998) 440--442.

\bibitem{LKF07}
J.~Leskovec, J.~Kleinberg, C.~Faloutsos, Graph evolution: Densification and
  shrinking diameters, ACM Trans. Knowl. Discov. Data 1~(1) (2007) 1--40.

\bibitem{KDD}
{KDD} {C}up '03, \url{http://www.cs.cornell.edu/projects/kddcup/} (2013).

\bibitem{CML11}
E.~Cho, S.~A. Myers, J.~Leskovec, Friendship and mobility: user movement in
  location-based social networks, in: Proceedings of the 17th ACM SIGKDD
  International Conference on Knowledge Discovery and Data Mining, ACM, 2011,
  pp. 1082--1090.

\bibitem{LLDM09}
J.~Leskovec, K.~J. Lang, A.~Dasgupta, M.~W. Mahoney, Community structure in
  large networks: Natural cluster sizes and the absence of large well-defined
  clusters, Internet Math. 6~(1) (2009) 29--123.

\bibitem{ML12}
J.~McAuley, J.~Leskovec, Learning to discover social circles in ego networks,
  in: Advances in Neural Information Processing Systems 25, 2012, pp. 548--556.

\bibitem{YL12}
J.~Yang, J.~Leskovec, Community-affiliation graph model for overlapping network
  community detection, in: Proceedings of the 12th International Conference on
  Data Mining, IEEE, 2012, pp. 1170--1175.

\bibitem{AJB99}
R.~Albert, H.~Jeong, A.-L. Barab{\'a}si, Internet: Diameter of the world-wide
  web, Nature 401~(6749) (1999) 130--131.

\bibitem{YL15}
J.~Yang, J.~Leskovec, Defining and evaluating network communities based on
  ground-truth, Knowledge and Information Systems 42~(1) (2015) 181--213.

\bibitem{BSB12}
N.~Blagus, L.~{\v{S}}ubelj, M.~Bajec, Self-similar scaling of density in
  complex real-world networks, Physica A 391~(8) (2012) 2794--2802.

\end{thebibliography}


\end{document}